\documentclass[twocolumn,showpacs,amsmath,amssymb]{revtex4}
\usepackage{graphicx}% Include figure files
\usepackage{dcolumn}% Align table columns on decimal point
\usepackage{bm}% bold math

\def\GGG{\frak G }
\def\gr3{\GGG\,(\SSS_3)}

\def\gr2{\GGG\,(\SSS_2)}

\def\SSS{\frak S}

\def\al{{\alpha}}
\def\bet{{\beta}}

\def\gam{{\gamma}}

\def\vp{\vspace}
\def\hp{\hspace}
\def\ed{\end{document}}
\def\beq{\begin{equation}}
\def\eeq{\end{equation}}
\def\bea{\begin{eqnarray}}
\def\eea{\end{eqnarray}}
\def\ba{\begin{array}}
\def\ea{\end{array}}
\def\bi{\begin{itemize}}
\def\ei{\end{itemize}}

\def\lb{\label}

\def\noi{\noindent}
\def\nn{\nonumber}

\numberwithin{equation}{section}

\begin{document}

%\arabic{section}

%\preprint{APS/123-QED}

\title{Path Integrals in Noncommutative Quantum Mechanics}% Force line breaks with \\

\author{Branko Dragovich}
\email{dragovich@phy.bg.ac.yu}
% \altaffiliation{Institute of Physics, P.O. Box 57,  11001 Belgrade, Serbia and Montenegro}
\affiliation{Institute of Physics, P.O. Box 57,  11001 Belgrade, Serbia and Montenegro}
 % \\
\author{Zoran Raki\'c }
\email{zrakic@matf.bg.ac.yu}
\affiliation{Faculty of Mathematics, P.O. Box 550, 11001 Belgrade,
Serbia and Montenegro}


\begin{abstract}
Extension of Feynman's path integral  to  quantum mechanics of
noncommuting spatial coordinates is considered. The corresponding
formalism for noncommutative classical dynamics related to
quadratic Lagrangians (Hamiltonians) is formulated. Our approach
is based on the fact that a quantum-mechanical system with a
noncommutative configuration space may be regarded as another
effective system with commuting spatial coordinates. Since path
integral for quadratic Lagrangians is exactly solvable and a
general formula for probability amplitude exists, we restricted
our research to this class of Lagrangians. We found general
relation between quadratic Lagrangians in their commutative and
noncommutative regimes. The corresponding noncommutative path
integral is presented. This method is illustrated by two
quantum-mechanical systems in the noncommutative plane: a particle
in a constant field and a harmonic oscillator.
%Valid PACS
%numbers may be entered using the \verb+\pacs{#1}+ command.
\end{abstract}

%Valid PACS appear here}% PACS, the Physics and Astronomy


%\pagestyle{plain}
\maketitle

% \arabic{section} \setcounter{section}{0}

%\sectionnumbering{arabic}
\section{Introduction}%    \label{sec:level1}First-level heading:\protect\\ The line
%break was forced \lowercase{via} \textbackslash\textbackslash}

 Already in the thirties of the last century, looking for an
approach to solve the problem of ultraviolet divergences,
Heisenberg and Schr\"odinger conjectured that spacetime
coordinates may be mutually noncommutative. However, the first
papers (see, e.g. \cite{snyder,moyal}) on this subject appeared at
the end of the forties and after some fifty years later an
increased interest in noncommutativity emerged in various quantum
theories. In particular, noncommutativity has been recently
intensively investigated in string/M-theory, quantum field theory
and quantum mechanics (for a review of noncommutative field theory
and some
related topics, see, e.g. \cite{nekrasov}). %\vp{1mm}

 Most of the research has been done in noncommutative field
theory, including its noncommutative extension of the Standard
Model \cite{wess}. Since quantum mechanics can be regarded as the
one-particle nonrelativistic sector of quantum field theory, it is
important to study its noncommutative aspects including connection
between ordinary and noncommutative regimes. In order to suggest
some possible phenomenological tests of noncommutativity,
noncommutative quantum mechanics (NCQM) of a charged particle in
the presence of a constant magnetic and electric fields has been
mainly considered on two- and three-dimensional spaces (see, e.g.
\cite{bellucci} and references therein). %\vp{1mm}

 Recall that  description of $D$-dimensional quantum-mechanical
system needs a Hilbert space $L_2 ({\mathbb R}^D)$ in which
observables are linear self-adjoint operators. In ordinary quantum
mechanics (OQM), classical canonical variables $x_k,p_j$ become
Hermitian operators $\hat{x_k},\hat{p_j}$  in $L_2 ({\mathbb
R}^D)$ satisfying the Heisenberg commutation relations \bea [\,
\hat{x_k},\hat{p_j}\,] = i\,\hbar\, \delta_{kj}, \
[\,\hat{x_k},\hat{x_j}\,] = 0,\ [\,\hat{p_k},\hat{p_j}\,]
=0,\label{e1.1} \eea where $k,j=1,2,\dots ,D$.

  In a very general NCQM  one maintains   $ [\, \hat{x_k},\hat{p_j}\,] =
i\,\hbar\, \delta_{kj}$, but some realization of $
[\,\hat{x_k},\hat{x_j}\,] \neq 0$ and $[\,\hat{p_k},\hat{p_j}\,]
\neq 0$ is   assumed.  We consider here the most simple and usual
NCQM which is based on the following algebra: \bea \label{e1.2}
\hp{-2mm} [\, \hat{x_k},\hat{p_j}\,] = i\,\hbar\, \delta_{kj}, \
\, [\,\hat{x_k},\hat{x_j}\,] = i\,\hbar\, \theta_{kj},\ \,
[\,\hat{p_k},\hat{p_j}\,] =0 , \eea where ${\Theta}=(\theta_{kj})$
is the antisymmetric matrix with constant elements. %\vspace{1mm}

  To find elements  $\Psi (x,t)$   of the Hilbert space in OQM,
 it is usually used the Schr\"odinger equation
\bea\label{e1.3}
i\,\hbar\, \frac{\partial}{\partial t}\, \Psi (x,t) =
\hat{H}\,\Psi (x,t),      \eea
 which realizes the eigenvalue problem for the
corresponding Hamiltonian operator $\hat{ H} = H (\hat{p}, x, t)$,
where\linebreak $\hat{p_k} =-\,i\,\hbar\,({\partial}/{\partial
x_k})$. %\vp{1mm}

 However, there is another approach based on the Feynman path
integral method \cite{feynman} \bea\label{e1.4} {\cal
K}(x'',t'';x',t') =\int_{(x')}^{(x'')} \exp \Big(
\frac{i}{\hbar}\, S[\,q\,]\Big)\, {\cal D}q , \eea where ${\cal
K}(x'',t'';x',t')$ is the kernel of the unitary evolution operator
$U (t)$ acting on $\Psi (x,t)$ in $L_2 ({\mathbb R}^D)$.

 Functional $ S[\,q\,] = \int_{t'}^{t''} L(\dot{q},q, t)\, dt$ is
the action for a path $q(t)$ in the classical Lagrangian
$L(\dot{q},q,t)$, and \linebreak $x''=q(t''), \ x'=q(t')$ are end
points with the notation $x =(x_1,x_2,\dots,x_n)$ and
$q=(q_1,q_2,\dots,q_n)$. The kernel ${\cal K}(x'',t'';x',t')$ is
also known as the probability amplitude for a quantum particle to
pass from position $x'$ at time  $t'$ to another point $x''$ at
$t''$, and is closely related to the quantum-mechanical propagator
and Green's  function. The integral in (\ref{e1.4}) has an
intuitive meaning that a quantum-mechanical particle may propagate
from $x'$ to $x''$ using infinitely many paths which connect these
two points and that one has to sum probability amplitudes over all
of them. Thus the Feynman path integral means a continual
(functional) summation of single transition amplitudes $\exp \left
( \frac{i}{\hbar}\, S[\,q\,]\right ) $ over all possible continual
paths $q(t)$ connecting $x'=q(t')$ and $x''=q(t'')$. In Feynman's
formulation, the path integral (1.4) is the limit of an ordinary
multiple integral over $N$ variables $q_i = q(t_i)$ when
$N\longrightarrow \infty$. Namely, the time interval $t''-t'$ is
divided into $N +1$ equal subintervals and integration is
performed for every $q_i \in (-\infty,+\infty)$ at fixed time
$t_i$.

 Path integral in its most general formulation contains
integration over paths in the phase space ${\Bbb R }^{2D} =
\{(p,q)  \}$ with fixed end points $x'$ and $x''$, and no
restrictions on the initial and final values of the momenta, i.e.
\bea \label{e1.5}& & \hspace{-7mm}
{\cal K}(x'',t'';x',t')=\vspace{2mm} \\
& & \hspace{-5mm} =  \int_{(x')}^{(x'')} \exp \Big( \frac{2\pi
i}{h} \int_{t'}^{t''}\big[\,p_k\, \dot{q}_k  - H(p,q,t)\,\big]\,
dt \Big)\, { {\cal D}q \, {\cal D}p}\, .\nn \eea However, for
Hamiltonians which are polynomials quadratic in momenta $p_k$, one
can explicitly perform integration over $-\infty < p_k < + \infty$
using the Gauss integrals and the corresponding path integral is
reduced to its form (\ref{e1.4}) (see Appendix).

 The Feynman path integral for quadratic Lagrangians  can be
evaluated analytically (see, e.g. a book \cite{steiner} and a
paper \cite{grosjean}) and the exact expression for the
probability amplitude is \bea \label{1.6} & & \hspace{-15mm} {\cal
K}(x'',t'';x',t') = \frac{1}{(i h)^{\frac{D}{2}}}
\sqrt{\det{\left(-\frac{\partial^2 {\bar S}}{\partial x''_k\,
\partial x'_j} \right)}} \vspace{2mm} \\ & & \hspace{12mm} \times\
\exp \bigg(\frac{2\,\pi\, i}{h}\,{\bar S}(x'',t'';x',t')\bigg),\nn
\eea where $ {\bar S}(x'',t'';x',t')$ is the action for the
classical trajectory which is the solution of the Euler-Lagrange
equation of motion.

 ${\cal K}(x'',t'';x',t')$, as the kernel of unitary evolution
operator, can be defined by equation \bea \Psi (x'',t'') = \int
{\cal K}(x'',t'';x',t')\, \Psi (x',t')\, dx' \label{e1.7} \eea and
then Feynman's path integral may be regarded as a method to
calculate this propagator. Eigenfunctions of the integral equation
(\ref{e1.7}) and of the above Schr\"odinger equation (\ref{e1.3})
are the same for the same physical system. Moreover, Feynman's
approach to quantum mechanics, based on classical Lagrangian
formalism, is equivalent to  Schr\"odinger's and Heisenberg's
quantization of the classical Hamiltonian. Note that the Feynman
path integral method, not only in quantum mechanics but also in
whole quantum theory, is intuitively more attractive and more
transparent in its connection with classical theory  than the
usual canonical operator formalism. In gauge theories, it is the
most suitable method of quantization. %\vp{.7mm}

 Besides its fundamental role in general formalism of quantum
theory, path integral method is an appropriate tool to compute
quantum phases. To this end, it has been partially considered  in
noncommutative plane for the Aharonov-Bohm effect
\cite{chaichian1}-\cite{turci} and a quantum system in a rotating
frame \cite{christiansen}. Our approach contains these and all
other possible systems with quadratic Lagrangians (Hamiltonians).
For three other approaches to the Feynman path integral in NCQM,
see Refs. \cite{mangano},  \cite{acatrinei} and \cite{smailagic}.
The difficulty of a straightforward generalization of the usual
path integral to NCQM is caused by an uncertainty of the end
position coordinates $x'$ and $x''$. These coordinates cannot be
exactly fixed, since $\Delta x'_k \, \Delta x'_j \geq
\frac{\hbar}{2}\, \theta_{kj} $ and $\Delta x''_k \, \Delta x''_j
\geq \frac{\hbar}{2}\, \theta_{kj}$ due to commutation relations
(\ref{e1.2}).

  In the Section II of this article we start by quadratic
Lagrangian which is related  to OQM and find the corresponding
quadratic Hamiltonian. Then we introduce quantum Hamiltonian with
noncommuting position coordinates and transform it to an effective
Hamiltonian with the corresponding commuting coordinates.   To
this end, let us note that algebra (\ref{e1.2}) of operators
$\hat{x_k}, \hat{p_j}$ can be replaced by the equivalent one
\bea\label{e1.8} [\, \hat{q_k},\hat{p_j}\,] = i\,\hbar\,
\delta_{kj}, \quad [\,\hat{q_k},\hat{q_j}\,] = 0,\quad
[\,\hat{p_k},\hat{p_j}\,] =0\, , \eea
 where linear transformation
\bea \hat{ x_k} = \hat{q_k} - \frac{\theta_{kj}\hat{p_j}}{2}
\label{e1.9} \eea is used, while $\hat p_k$ are retained
unchanged, and summation over repeated indices $j$ is assumed.
Owing to the transformation (\ref{e1.9}), NCQM related to  the
classical phase space with points $(p,x)$ can be regarded as an
OQM on the other phase space $(p,q)$. Thus, in $q$-representation,
$\hat{p_k} = -\,i\,\hbar\, ({\partial}/{\partial q_k})$ in  the
equations (\ref{e1.7}) and (\ref{e1.8}). An alternative way to
find the commutative analog of a quantum-mechanical system on
noncommutative space is to use the Moyal star product \cite{moyal}
in the Schr\"odinger equation. Such treatment  is equivalent to
the change of potential $V({\hat x})$ by using transformation
(\ref{e1.9}) \cite{mezincescu}. It is worth noting that
Hamiltonians\linebreak $H(\hat{p},\hat{x}, t)
=H(-i\hbar\,({\partial}/{\partial q_k}),\, q_k +
({i\hbar\,\theta_{kj}}/{2})\,({\partial}/{\partial q_j}),\, t )$,
which are more than quadratic in $\hat{x}$, will induce
Schr\"odinger equations with derivatives higher than second order
and even of the infinite order. This leads to a new part of modern
mathematical physics of partial differential equations with
arbitrary higher-order derivatives. In this paper we restrict our
consideration to the case of  Lagrangians quadratic in $q$ and
$\dot{q}$. Transforming new Hamiltonian to Lagrangian we find an
effective Lagrangian related to the system with noncommuting
spatial coordinates.

In Sec. III we give a general formula for the path integral
expressed in terms of the classical action for  the noncommutative
quadratic Lagrangians. Summary and a short discussion are
presented in Sec. IV. In Appendix we show the transition from the
phase space path integral to the usual Feynman coordinate one for
the general case of quadratic Hamiltonians.

\section{Classical Dynamics on Noncommutative Spaces}

\subsection{Connection between Lagrangian and Hamiltonian in
commutative regime}

 Let us start with a classical system described by a quadratic
Lagrangian which the most general form in three dimensions is:
\bea \nn \hskip-5mm  & & \hskip-7mm L(\dot{x}, x,t)  \\
\nn &=& \alpha _{11}\,\dot{x}_1^2 +\alpha _{12}\,\dot x_1 \, \dot
x_2 +\alpha _{13}\,\dot x_1 \, \dot x_3 \, +\alpha
_{22}\,\dot{x}_2^2
  \vp{2mm} \\ \nn & + & \alpha _{23}\,\dot x_2 \, \dot x_3 \, + \alpha _{33}\,\dot{ x}_3^2 + \beta
_{11}\,{\dot x}_1 \, x_1  +\beta _{12}\, \dot x_1 \, x_2 \label{8}  \vp{2mm}\\
\nn & + &  \beta _{22}\,\dot x_2 \, x_2 +\beta _{23}\,\dot x_2 \,
x_3 + \beta _{31}\,\dot x_3\, x_1  +\beta _{32}\, \dot x_3
\, x_2 \vp{2mm}\\
 & + &  \beta _{33}\, \dot x_3 \, x_3 + \beta
_{13}\,\dot x_1 \, x_3 + \beta _{21}\,\dot x_2 \, x_1 +\gamma
_{11}\, x_1^{2}  \vp{2mm} \nn \\
& + & \gamma _{12}\, x_1 \, x_2 +\gamma _{13}\, x_1 \, x_3  +
\gamma _{22} \, x_2 ^{2} + \gamma _{23}\, x_2 \, x_3 \nn \vp{2mm}
\\ & + & \gamma _{33}\, x_3^{2} + \delta _1 \, \dot x_1  +\delta _2 \,
\dot x_2 + \delta_3 \, \dot x_3 + \eta_1 \, x_1 \nn \vp{2mm}
\\ & + & \eta _2\, x_2
+\eta_3\, x_3 + \phi , \label{e2.1} \eea
 where the coefficients $\alpha_{ij}
=\alpha_{ij}(t),\ \beta_{ij} =\beta_{ij}(t),\ \gamma_{ij}
=\gamma_{ij}(t),$  $ \delta_{i} =\delta_{i}(t),$ \ $\eta_{i}
=\eta_{i}(t)$ and $\phi =\phi(t)$ are some analytic functions of
the time $t$ .

 If we introduce the  matrices
\bea && \hp{-3mm}\begin{array}{ccc}  {\alpha } = \left(
 \begin{array}{ccc}
 \alpha _{11} & {\alpha _{12} \over 2} &
{\alpha _{13} \over 2}\vspace{2mm} \\
{\alpha_{12} \over 2} & \alpha_{22} & {\alpha_{23} \over
2} \vspace{2mm}\\
 {\alpha_{13} \over 2 } & {\alpha_{23} \over 2} &
 \alpha_{33}
\end{array}
\right) ,   &   {\beta } = \left(
\begin{array}{ccc}
 \beta_{11} & \beta_{12} & \beta_{13} \\
 \beta_{21} & \beta_{22} & \beta_{23} \\
 \beta_{31} & \beta_{32} & \beta_{33} \end{array}
\right) ,  &  \vspace{3mm}\end{array}\nn \\ &&  \label{e2.2} \\
\nn && \hp{-3mm}  \begin{array}{ccc} {\gamma }= \left(
\begin{array}{ccc} \gamma_{11} & {\gamma_{12} \over 2}
& {\gamma_{13} \over 2}\vspace{2mm} \\
{\gamma_{12} \over 2}  & \gamma_{22} & {\gamma_{23}\over 2}\vspace{2mm} \\
{\gamma_{13} \over 2 } & {\gamma_{23} \over 2} & \gamma_{33}
\end{array} \right) ,   & \begin{array}{l} {\delta }^{\tau} =(\delta _1 ,\delta_2
,\delta_3 ) ,\vspace{2mm} \\  {\eta}^{\tau} = (\eta_1 , \eta_2 ,
\eta_3 ) ,\end{array} \end{array} \eea \vspace{2mm}
 and assuming that the matrix $\al$ is nonsingular (regular), and
vectors \bea\label{e2.3}
 {\dot x}^{\tau}= (\dot x_1, \dot x_2, \dot x_3), \quad {x}^{\tau}= (x_1 ,x _2,x_3),
\eea then one can express the Lagrangian (\ref{e2.1}) in the
following, more compact, form: \bea \nn   \hp{-25mm}
L(\dot{x},x,t) & = & \langle\, {\alpha }\, {\dot x }, {\dot x }
\,\rangle + \langle\, {\beta }\, {x }, {\dot x }\,
\textbf{}\rangle +\langle\, {\bf \gamma }\, {x}, {x }\, \rangle
\vspace{2mm}
\\  & & +\    \langle\, {\delta } , {\dot x }\, \rangle +
\langle\, {\eta } , {x }\, \rangle + \bf \phi  ,\label{e2.4} \eea
where index $\tau$ imposes transpose map and $\langle\, \cdot\,
,\, \cdot\, \rangle$ denotes standard scalar product. Using the
equations
\begin{equation}\label{e2.5}
p_j = {\partial L \over \partial \dot x_j}, \ \ \ j=1,2,3,
\end{equation}
one can express ${\dot x }$ as
\begin{equation}\label{e2.6}
{\dot x } = {1\over 2}\,\, {\alpha^{-1}}\, ({p}  - {\beta}\, {x }
- \delta ).
\end{equation}
Then the corresponding classical  Hamiltonian \beq \label{e2.7}
H(p,x,t)=\langle\, {p}, \dot {x }\, \rangle - L(\dot{x},x,t)\eeq
becomes also quadratic, i.e. \bea \nn\label{e2.8} \hp{-15mm}
H(p,x,t) & = & \langle\, {A}\, {p}, {p}\, \rangle + \langle\,
{B}\, {x}, {p }\, \rangle +\langle\, {C}\, {x }, {x}\, \rangle
\vspace{2mm} \\ & & +\  \langle\, {D } , {p} \,\rangle + \langle\,
{ E } , {x} \rangle + F , \eea where: \bea
  && \hp{-10mm} {A } ={\frac 14}\,\,  {\alpha}^{-1}, \hspace{1.85cm}   {B }
=-\,{1\over 2} \,\, {\alpha}^{-1}\, {\beta}, \vp{2mm}\nn \\
& & \hp{-10mm} {C} ={1\over 4}\,\, {\beta}^{\tau}\,
{\alpha}^{-1}\, {\beta} - {\gamma } , \hspace{.5cm}
  {D} =- {1\over 2}\,\, {\alpha }^{-1}\,  \delta,\vp{2mm} \label{e2.9}\\
& & \hp{-10mm} {E } = {1\over 2}\,\, {\beta}^{\tau}
{\alpha}^{-1}\, {\delta} - {\eta} , \hspace{.69cm} {F} = {1\over
4}\,\, {\delta}^{\tau}\, {\alpha}^{-1}\, {\delta}  - {\phi } \nn
\,. \eea  %\vp{1mm}

  Let us note now that matrices $A$ and $C$ are symmetric
($A^{\tau} = A$ and $C^{\tau} = C$), since the matrices $\al$ and
$\gam$ are symmetric. If the Lagrangian $L(\dot{x},x,t)$ is
nonsingular ($\det {\al}\neq 0$) then the Hamiltonian $H(p,x,t)$
is also nonsingular ($\det { A}\neq 0$). %\vspace{2mm}

 The above calculations can be considered as a map from the space
of quadratic nonsingular Lagrangians ${\cal L}$ to the
corresponding space of quadratic nonsingular Hamiltonians ${\cal
H}\,.$ More precisely, we have $ \varphi: {\cal L}\longrightarrow
{\cal H}, $ given by \bea \nn
& & \hspace{-15mm} \varphi(L({\al},{\bet},{\gam},{\delta},{\eta} ,{\phi},{\dot x},x)) \vspace{3mm} \\
&  & \hspace{-12mm} = H (\varphi_1(L),\varphi_2(L),
\varphi_3(L),\varphi_4(L),\varphi_5(L),\varphi_6(L),
\vspace{4mm}\nn\\ &  & \hspace{-8mm} \varphi_7(L),\varphi_8(L)) =
H({A},{B},{C},{D},{E},{F},{p},{x})\,.\lb{e2.10}\eea From relation
(\ref{e2.10}) it is clear that inverse of $\varphi$ is given by
the same relations (\ref{e2.9}). This fact implies that $\varphi$
is essentially an involution, i.e. $\varphi\circ\varphi= id\,.$

\subsection{ Connection between Lagrangian and Hamiltonian in
noncommutative regime }

 In the case of noncommutative spatial coordinates $[\,\hat x_k ,
\hat x_j \, ] = i\, \hbar\,\theta _{kj}$, one can replace these
coordinates using the following ansatz  \beq\label{e2.11} {{\hat
x}}= {{\hat q}} - \frac12\,\, {\Theta}\, {{\hat p}}\,,\eeq where,
as usually, ${{\hat x}}^{\tau}=(\hat x_1,\hat x_2,\hat x_3),
\,{{\hat q}}^{\tau}=(\hat q_1,\hat q_2,\hat q_3 ),\,$ and  $
{{\hat p}}^{\tau}=(\hat p_1,\hat p_2,\hat p_3 )$ are operators and
\bea\nn { \Theta} = \left(
\begin{array}{ccc}
0             & \theta _{12}  & \theta _{13}\\
-\,\theta _{12} & 0             & \theta _{23} \\
-\,\theta _{13} & -\,\theta _{23} & 0 \end{array} \right) . \eea
Now, one can easily check that $\hat q_i$ for $i = \ 1,2,3 \ $ are
mutually commutative operators (but do not commute with operators
of momenta, i.e. $[\,\hat q_k , \hat p _j\, ] = i\,\hbar\, \delta
_{kj}$). %\vspace{1mm}

If we start with quantization of the nonsingular quadratic
Hamiltonian given by the relation (\ref{e2.8}), i.e. \linebreak
$\hat H={ H}({A},{B},{C},{D},{E},{F},{{\hat p}}, {\hat x})$ and
then apply the change of coordinates (\ref{e2.11}), we will again
obtain quadratic quantum Hamiltonian, ${\hat H}_{\theta}\, =
H_\theta (\hat{p}, \hat{q}, t)$:  \bea \hp{-15mm}\hat H_{\theta} &
= & \langle\, {A}_{\theta}\, \hat {p}, \hat {p}\, \rangle +
\langle\, {B}_{\theta}\, \hat {q}, \hat {p}\, \rangle +\langle\,
{C}_{\theta}\, \hat {q}, \hat {q}\, \rangle \vspace{2mm}\nn \\ & &
+\ \langle\, {D}_{\theta} , \hat {p}\, \rangle + \langle\,
{E}_{\theta} , \hat {q}\, \rangle + {F}_{\theta} , \label{e2.12}
\eea where \bea\nn & & \hspace{-12mm} {A}_{\theta} = ({A} -
{1\over 2}\,\, {B}\, { \Theta} -{1\over 4}\, {\Theta}\, {C}\, {
\Theta})_{sym}, \ \
{B}_{\theta} = {B} + {\Theta}\, {C} , \vp{2mm}\\
& &  \hspace{-12mm} {C}_{\theta} = {C}, \ \  {D}_{\theta} = {D} + {1\over
2}\,\, {\Theta}\, {E},\ \  {E}_{\theta} = {E}, \ \
{F}_{\theta} = {F} ,\label{e2.13} \eea and  $_{sym}$ denotes
symmetrization of the corresponding operator. Let us note that for
the nonsingular Hamiltonian $\hat H$ and for  sufficiently small
$\theta_{kj}$ the Hamiltonian $\hat H_{\theta}$ is also
nonsingular.

  In the process of calculating path integrals for the above
systems, we need classical Lagrangians. It is clear that to an
arbitrary quadratic quantum Hamiltonian we can  associate the
classical one replacing operators by the corresponding classical
variables. Then, by using equations $$ \dot{q}_k = \frac{\partial
H_\theta}{\partial p_k} , \ \ \ k=1,2,3,
$$ from such Hamiltonian we can come back to the
corresponding Lagrangian $$L_\theta (\dot{q},q,t) = \langle\, p,
\dot{q} \, \rangle - H_\theta (p,q,t),
$$ where
$$ p = \frac{1}{2} \, A_\theta^{-1}\, (\dot{q} - B_\theta\, q - D_\theta) $$
is replaced in $H_\theta (p,q,t)$. In fact, our idea is to find
connection between Lagrangians of  noncommutative and the
corresponding commutative quantum mechanical systems (with $\theta
= 0 $). This implies to find the composition of the following
three maps: \beq \label{e2.14}
L_{\theta}=(\varphi\circ\psi\circ\varphi)(L), \eeq where
$L_{\theta}=\varphi(H_{\theta}),\, H_{\theta}=\psi(H)$ and
$H=\varphi(L)\,$ (here we use facts that $\varphi$ is an
involution given by formulas (\ref{e2.9}), and $\psi$ is given by
(\ref{e2.13})). More precisely, if \bea \nn \hp{-15mm}
L(\dot{x},x,t) &=&\langle\, {\alpha}\, {\dot x}, {\dot x}
\,\rangle + \langle\, {\beta}\, {x}, {\dot x}\, \rangle +\langle\,
{\gamma}\, {x}, {x }\, \rangle \vspace{2mm} \nn \\ & & +\
\langle\, {\delta} , {\dot x }\, \rangle + \langle\, {\eta} ,
{x}\, \rangle + \phi ,\label{e2.15}\eea      and \bea \hp{-10mm}
L_{\theta} (\dot{q},q,t) &=& \langle\, {\alpha}_{\theta}\, {\dot
q}, {\dot q} \,\rangle + \langle\, {\beta}_{\theta}\, {q}, {\dot
q}\, \rangle +\langle\, {\gamma}_{\theta}\, {q}, {q }\,
\rangle\vspace{2mm} \nn
\\ & & +\ \langle\, {\delta}_{\theta} , {\dot q }\, \rangle +
\langle\, {\eta}_{\theta} , {q} \, \rangle + {\phi}_{\theta}\,
, \label{e2.16} \eea then the connection between their coefficients is given
 by \bea\nn   {\al}_{\theta} &
=& \big[\, {\al}^{-\,1} - \frac 12 \,\, (\Theta\,
{\bet}^{\,\tau}\, {\al}^{-\,1}-  {\al}^{-\,1}\, {\bet}\, \Theta)
\vspace{2mm} \nn \\ & & +\ \Theta\, {\gam}\, \Theta - \frac 14\,\,
\Theta\, {\bet}^{\tau}\, {\al }^{-\,1}\, {\bet}
\,\Theta\, \big]^{-\,1}\,, \nn \vp{2mm} \\
 \nn   {\bet}_{\theta} &= & {\al}_{\theta}\, \big( {\al}^{-\,1}\,
 {\bet} - \frac 12 \,\, \Theta\, {\bet}^{\,\tau}\, {\al}^{-\,1}\,
{\bet} + 2\, \Theta\, {\gam} \big)\,, \vp{2mm} \\
\nn  {\gam}_{\theta} &=&\frac 14\,\, \big({\bet}^{\,\tau}\,
{\al}^{-\,1}+\frac 12\,\, {\bet}^{\,\tau}\, {\al}^{-\,1}\,
\bet\,\Theta\, - 2\,  {\gam}\, \Theta \big)\, \vspace{2mm} \nn \\
& & \times\ {\al}_{\theta}\, \big({\al}^{-\,1}\,{\bet}- \frac
12\,\, \Theta\, {\bet}^{\,\tau}\, {\al}^{-\,1}\, {\bet} +\ 2\,
\Theta\, {\gam}\big) \vspace{2mm} \nn \\ & & \nn - \ \frac 14\,\,
{\bet
}^{\,\tau}\, {\al}^{-\,1}\, {\bet} +{\gam}\,,\nn\vp{2mm} \\
{\delta}_{\theta}& =&
{\al}_{\theta}\,\big({\al}^{-\,1}\,{\delta}-\frac
12\,\,\Theta\,{\bet}^{\,\tau}\,
{\al}^{-\,1}\,{\delta} +\Theta\,\eta \big)\,, \nn \vp{2mm} \\
{\eta }_{\theta}& = & \frac 12\, \, \big(\bet^{\,\tau}\,
{\al}^{-\,1}+\frac 12\, {\bet}^{\,\tau}\, {\al}^{-\,1}\,
\bet\,\Theta\, - 2\,  {\gam}\, \Theta \big)\, \vspace{2mm} \nn \\
& & \times\ {\al}_{\theta}\, \big({\al}^{-\,1}\,{\delta} -\ \frac
12\,\, \Theta\, {\bet}^{\,\tau}\, {\al}^{-\,1}\,{\delta} +
\Theta\, {\eta}\big) \vspace{2mm} \nn \\ & &  - \ \frac 12\,\,
{\bet}^{\,\tau}\,{\al}^{-\,1}\, {\delta} + {\eta}\,,\nn \vp{2mm} \\
 {\phi }_{\theta} & = & \frac 14\,\, \langle\,
{\delta}_{\theta}\,, {\al}^{-\,1}\,{\delta}-\frac
12\,\,\Theta\,{\bet}^{\,\tau}\, {\al}^{-\,1}\,{\delta}
+\Theta\,{\eta}\,\rangle \vspace{2mm} \nn \\ & &  -\ \frac 14\,\, \langle\,
{\al}^{-\,1}\,{\delta}\,,{\delta}\,\rangle + {\phi}\,. \label{e2.17} \eea

  It is clear that formulas (\ref{e2.17}) are very complicated and
that to find explicit  exact relations between elements of
matrices in general case is a very hard task. However, the
relations (\ref{e2.17}) are quite useful in all particular cases.
On the above classical analog of Hamiltonian  (\ref{e2.12}) and
Lagrangian (\ref{e2.16}) one can apply usual techniques of
classical mechanics.

\section{Noncommutative Path Integrals}

Although derived for three-dimensional space, the results obtained
in the preceding section are valid for an arbitrary spatial
dimensionality D. If we know Lagrangian (\ref{e2.15}) and algebra
(\ref{e1.2}) we can obtain the corresponding effective Lagrangian
(\ref{e2.16}) suitable for quantization with path integral in
NCQM. Exploiting the Euler-Lagrange equations \bea  \frac{\partial
L_\theta}{\partial q_k} -\frac{d}{dt}  \frac{\partial L_\theta}
{\partial{\dot q}_k} =0 ,   \quad k=1,2,\cdots, D  \nn \eea one
can obtain classical path $q_k =q_k (t)$ connecting given end
points $x' = q(t')$ and  $x''= q(t'')$. For this classical
trajectory one can calculate action ${\bar S}_\theta
(x'',t'';x',t') =\int_{t'}^{t''} L_\theta (\dot{q}, q,t)\, dt$.
Path integral in NCQM is a direct analog of (\ref{e1.4}) and its
exact expression in the form of quadratic actions ${\bar
S_{\theta}}(x'',t'';x',t')$ is \bea & & \hspace{-12mm} {\cal
K}_{\theta}(x'',t'';x',t') = \frac{1}{(i h)^{\frac{D}{2}}}
\sqrt{\det{\left(-\frac{\partial^2 {\bar S_{\theta}}}{\partial
x''_k\,
\partial x'_j} \right)}} \vspace{2mm} \label{e3.1} \\ & & \hspace{12
mm} \times\  \exp \left(\frac{2\,\pi\, i}{h}\,{\bar
S_{\theta}}(x'',t'';x',t')\right)\, . \nn \eea

 We present here two examples for $D=2$ case.

 \subsection{A particle in a constant field}

 In this example the  Lagrangian on commutative configuration
space is \bea\label{e3.2} L(\dot{x},x) = \frac{m}{2}\,
(\dot{x}_1^2 + \dot{x}_2^2 ) -\eta_1\, x_1  - \eta_2\, x_2 . \eea

 The corresponding  data in the matrix form are: \beq\label{e3.3}
  \ba{l} \displaystyle{{\al}=\frac m2\,\, { I
 },\quad
{\bet}=0,\quad {\gam}=0, \quad {\delta}=0,} \vp{3mm}  \\
  \displaystyle{{\eta}^\tau =(-\eta_1^{},-\eta_2^{}),\quad {\phi}=0,}
  \ea\eeq where
${ I}$ is $2\times 2$ unit matrix. %\vspace{2mm}

 Using the general composition formula (\ref{e2.17}), one can
easily find \beq   \ba{l} \displaystyle{{\al}_{\theta}=\frac
{m}{2}\,\, { I},\quad
 {\bet}_{\theta}=0, \quad  {\gam}_{\theta}=0, \quad
 {\eta}_{\theta}={\eta},} \vp{3mm} \\
 \displaystyle{{\delta}_{\theta}^\tau= \frac{m\,\theta}{2}\, (-\eta_2 ,
\eta_1)\,, \quad
 \phi_{\theta}=\frac{m\,\theta^2}{8}\,(\eta_1^2 +
\eta_2^2)\,.}  \label{3.4}\ea\eeq

  In this case, it is easy to find the classical action. The
Lagrangian $L_{\theta}(\dot{q},q,t)$ is \bea\lb{e17c} L_{\theta} &
= & \frac{m}{2} \, \, ( {\dot{q}_1}^2 +{\dot{q}_2}^2 ) + \frac{m\,
\theta}{2} \,(  \eta_1\,
\dot{q}_2 - \eta_2\, \dot{q}_1 ) - \eta_1\, q_1 \vspace{2mm} \nn \\
 &  & - \ \eta_2\, q_2
+\frac{m\,\theta^2}{8}\,({\eta_1}^2+{\eta_2}^2)\,. \label{e3.5}
\eea The Lagrangian given by (\ref{e3.5}) implies the
Euler-Lagrange equations \beq\lb{e3.6} m\, \ddot{q}_1 = -\eta_1,
\qquad m\, \ddot{q}_2 = -\eta_2\,. \eeq %\vp{0mm}

 Their solutions are: \beq\label{e3.7} \hp{-12mm}  \ba{l}
\displaystyle{q_1(t)= -
\frac{\eta_1\,t^2}{2\,m}+t\,C_2+C_1,}\vspace{2mm}\\
\displaystyle{q_2(t)=
-\frac{\eta_2\,t^2}{2\,m}+t\,D_2+D_1,}\ea\eeq where $C_1,C_2,D_1$
and $D_2$ are constants which have to be determined from
conditions:
 \bea\lb{e3.8} \hp{-8mm} q_1(0)=x_1',\, q_1(T)=x_1'', \,
q_2(0)=x_2',\,  q_2(T)=x_2''.\eea  After finding the corresponding
constants, we  have
\beq\label{e3.9} \hp{-2mm}  \ba{l} \displaystyle{q_j(t)= x_j'-
\frac{\eta_j\,t^2}{2\,m}+ t\,\left(\frac{1}{T}\,(x_j''-x_j') +
\frac{\eta_j\, T}{2\,m}\right),}
  \vspace{2mm} \\
\displaystyle{\dot{q}_j(t)= - \frac{\eta_j\,t}{m}+
\frac{1}{T}\,(x_j''-x_j') + \frac{\eta_j\, T}{2\,m}\,,\ j=1,2\,.}
\ea\eeq

Using (\ref{e3.8}) and (\ref{e3.9}),  we finally calculate the
corresponding action \bea & & \hspace{-5mm} {\bar S}_\theta
(x'',T;x',0)=\int\limits_{0}^T \, L_{\theta}(\dot{q},q,t)\, d\,t
\vp{3mm}\nn \\ & & \hp{-3mm} = \ \nn \frac{m}{2\,T} \,
\big[(x_1''-x_1')^2+(x_2''-x_2')^2\big]- \frac{T}{2}\,\,
\big[\eta_1\,(x_1''+
x_1')  \vp{3mm} \\
& & \hp{-3mm}+\ \eta_2\,(x_2''+x_2')\big] \nn  +
\frac{m\,\theta}2\,\,
\big[\eta_1\,(x_2''- x_2') - \eta_2\,(x_1''- x_1')\big] \vp{3mm} \\
& & \hp{-3mm}-\  \frac{T^3}{24\, m}\,\, ({\eta_1}^2+{\eta_2}^2) +
\frac{m\,\theta^2\,T}{8}\,\, ({\eta_1}^2+{\eta_2}^2)\,.
\label{e3.10}\eea

 According to (\ref{e3.1}) one gets \bea\nn
 & & \hspace{-10mm} {\cal K}_\theta (x'',T;x',0) =\frac{1}{ih} \frac{m}{T} \exp\left(
 \frac{2\pi i}{h}\, \bar{S}_\theta (x'',T;x',0)   \right)
\vspace{2mm}\nn \\
& & \hspace{-8mm} =\ {\cal K}_0 (x'',T;x',0)\, \exp\left(
\frac{2\pi i}{h}\frac{m\,\theta}{2} \big[\eta_1 (x''_2 -x'_2) \right.\vspace{2mm}\nn \\
& & \hspace{-5mm}- \ \left.\eta_2 (x''_1 -x'_1)
+\frac{\theta\,T}{4}\, (\eta_1^2 +\eta_2^2)\big]
\right),\label{e3.11} \eea where ${\cal K}_0 (x'',T;x',0)$ is related
to the Lagrangian (\ref{e3.2}) for which $\theta =0$. Hence, in this case there is a
difference only in the phase factor.

 It is easy to see that the following connection holds: \bea
\hspace{-3mm} {\cal K}_\theta (x'',T;x',0) =\ {\cal K}_0
(x''+\frac{\theta\,T}{2}\,J\eta,T;x',0) ,\label{e3.12} \eea where
\bea  J=\left( \ba{rc} 0 & 1\\
-\,1 & 0\ea \right) . \nn
\eea

 \subsection{ Harmonic oscillator}

 The commutative Lagrangian in the question is \bea\lb{e3.13}
\hp{-5mm} L(\dot{x},x) = \frac{m}{2}\, (\dot{x}_1^2  +
\dot{x}_2^2) -\frac{m\,\omega^2}{2}\, (x_1^2 + x_2^2)\,. \eea

  In this case we have \bea\nn \hspace{-1.2cm}
{\al}=\frac m2\,\, { I},\ \   {\bet}=0,\quad
{\gam}=-\,\frac{m\,\omega^2}{2}\,{ I},\ \Theta = \theta\, J,  \eea
\bea  & & \hp{-10mm} \ \ J^2=-\, { I},\quad {\delta}=\eta=0,\quad
\phi= 0 \label{e3.14} . \eea

  Using formulas (\ref{e2.17}), one can easily find \bea & &
\hp{-10mm} {\al}_{\theta}=\frac {m}{2\, \kappa}\, { I},\quad
 \label{e3.20} {\bet}_{\theta} =\frac {m^2\,\omega^2\,\theta}{2\,\kappa}\,
 J^{\tau},\quad {\gam}_{\theta}=- \frac {m\, \omega^2}{2\, \kappa}\,
{ I}\,,\vp{2mm}\nn \\ & & \hp{-10mm}
\delta_{\theta}=\delta_{\theta}=0,\quad \phi_{\theta}=0,\eea where
$\kappa = 1+\frac{m^2\, \omega^2\, \theta^2}{4}.$

 The  corresponding noncommutative Lagrangian is \bea\lb{e3.16} &
& \hspace{-5mm} L_{\theta}( \dot{q},q)  =
 \nn \ \frac{m}{2\, \kappa} \,  \big(
{\dot{q}_1}^2 +{\dot{q}_2}^2\big)+
\frac{m^2\,\omega^2\,\theta}{2\, \kappa} \, \big( \dot{q}_2\,q_1
- \dot{q}_1\, q_2\big)\vspace{2mm} \\ & & \hspace{15mm}
     -\  \frac{m\,\omega^2}{2\, \kappa}\, \big({q_1}^2 +{q_2}^2\big) \,. \eea
From (\ref{e3.16}), we obtain the  Euler-Lagrange equations,
\beq\lb{e3.17} \ba{l} \displaystyle{\ddot{q}_1-
m\,\omega^2\,\theta\,\dot{q}_2+ \omega^2\,q_1 =0,} \vspace{2mm} \\
\displaystyle{\ddot{q}_2+ m\,\omega^2\,\theta\,\dot{q}_1+
\omega^2\,q_2 =0\,.} \ea\eeq Let us remark that the Euler-Lagrange
equations (\ref{e3.17}) form a coupled system of second order
differential equations, which is more complicated than in
commutative case ($\theta=0$). One can transform the system
(\ref{e3.17})  to  \beq\lb{e3.18} \ba{l}
\displaystyle{{q}_1^{(4)}+ \omega^2\,(2+
m^2\, \omega^2\, \theta^2)\,{q}_1^{(2)}+ \omega^4\,q_1 =0,} \vspace{2mm}\\
 {q}_2^{(4)}+ \omega^2\,(2+ m^2\, \omega^2\,
\displaystyle{\theta^2)\,{q}_2^{(2)}+ \omega^4\,q_2 =0\,.} \ea\eeq
The solution of the equations (\ref{e3.17}) has the following form
\beq\lb{e3.19} \ba{l} \displaystyle{{q}_1(t) = C_1 \cos ( y_1t) +
C_2 \sin (y_1 t)} \vp{1mm}
\\  \hp{12mm}  \displaystyle{+\ C_3 \cos (y_2t)
+ C_4 \sin (y_2t)\,,}  \vp{2mm}\\
\displaystyle{{q}_2(t) = D_1 \cos (y_1t) + D_2 \sin (y_1t)}\vp{1mm}\\
 \hp{12mm} \displaystyle{+\ D_3 \cos (y_2t) + D_4 \sin (y_2t)\,,} \ea\eeq
where ($\omega>0$)\,   \bea\nn \ba{l} \displaystyle{{y}_1 =
\frac{m\, \theta\, \omega^2
+\omega\,\sqrt{4+m^2\,\theta^2\,\omega^2}}{2}=\frac{m\, \theta\,
\omega^2
+2\,\omega\,\sqrt{\kappa}}{2}} \vspace{3mm} \\
\displaystyle{{y}_2 = \frac{m\, \theta\, \omega^2
-\omega\,\sqrt{4+m^2\,\theta^2\,\omega^2}}{2}=\frac{m\, \theta\,
\omega^2 - 2\, \omega\,\sqrt{\kappa}}{2} \ .} \ea\eea If we impose
connections between $q_1$ and $q_2$ given by (\ref{e3.17}), we
obtain the following connection between constants $C$ and $D$:
\bea\lb{e3.20} \hp{-8mm} D_1=C_2,\ \ D_2=-C_1,\ \ D_3=C_4,\ \
D_4=-C_3\,.\eea The unknown constants $C_1,C_2,C_3$ and $C_4$ one
can find from initial conditions given by (\ref{e3.8}). Then one
finally obtains the solutions \bea\nn {q}_1 &=& \frac{1}{2}
\Big(\Big(x_1'+x_2'\, \cot (\omega\,
\sqrt{\kappa }\,\,T)- \csc (\omega\, \sqrt{\kappa }\,\,T) \\
\nn  &\times& \big( x_2''\cos{\frac{m\,\theta\,\omega^2\,T}{2}}+
x_1''\sin{\frac{m\,\theta\,\omega^2\,T}{2}}\big)\Big)\cos{(y_1t)}
\\ \nn &+& \Big(x_2' -x_1'\, \cot (\omega\, \sqrt{\kappa }\,\,T)+ \csc
(\omega\, \sqrt{\kappa }\,\,T)\\ \nn  &\times & \big(
x_1''\cos{\frac{m\,\theta\,\omega^2\,T}{2}}  -
x_2''\sin{\frac{m\,\theta\,\omega^2\,T}{2}}\big)\Big)\sin{(y_1t)}\\
\nn &+& \Big(x_1'-x_2'\, \cot (\omega\, \sqrt{\kappa }\,\,T) +
\csc (\omega\, \sqrt{\kappa }\,\,T) \\ \nn &\times & \big(
x_2''\cos{\frac{m\,\theta\,\omega^2\,T}{2}}+
x_1''\sin{\frac{m\,\theta\,\omega^2\,T}{2}}\big)\Big) \cos{(y_2t)
}\\
\nn &+& \Big(x_2'+x_1'\, \cot (\omega\,
\sqrt{\kappa }\,\,T)+ \csc (\omega\, \sqrt{\kappa }\,\,T) \\
 \nn  &\times& \big(- x_1''\cos{\frac{m\,\theta\,\omega^2\,T}{2}}+
x_2''\sin{\frac{m\,\theta\,\omega^2\,T}{2}}\big)\Big)\sin{(y_2t)}
\Big)\, ,
 \vspace{2mm} \\ \nn {q}_2 &=& \frac{1}{2}
\Big(\Big(x_2'-x_1'\, \cot (\omega\,
\sqrt{\kappa }\,\,T)- \csc (\omega\, \sqrt{\kappa }\,\,T) \\
 \nn &\times& \big(
x_1''\cos{\frac{m\,\theta\,\omega^2\,T}{2}}-
x_2''\sin{\frac{m\,\theta\,\omega^2\,T}{2}}\big)\Big)\cos{(y_1t)}
\\ \nn &-& \Big(x_1' +x_2'\, \cot (\omega\, \sqrt{\kappa }\,\,T)- \csc
(\omega\, \sqrt{\kappa }\,\,T)\\ \nn &\times& \big(
x_2''\cos{\frac{m\,\theta\,\omega^2\,T}{2}}  +
x_1''\sin{\frac{m\,\theta\,\omega^2\,T}{2}}\big)\Big)\sin{(y_1t )}\\
\nn &+& \Big(x_2'+x_1'\, \cot (\omega\, \sqrt{\kappa }\,\,T) +
\csc (\omega\, \sqrt{\kappa }\,\,T) \\ \nn &\times&  \big(
-x_1''\cos{\frac{m\,\theta\,\omega^2\,T}{2}}+
x_2''\sin{\frac{m\,\theta\,\omega^2\,T}{2}}\big)\Big) \cos{(y_2t)}\\
\nn &-& \Big(x_1'-x_2'\, \cot (\omega\,
\sqrt{\kappa }\,\,T)+ \csc (\omega\, \sqrt{\kappa }\,\,T) \\
 &\times& \big( x_2''\cos{\frac{m\,\theta\,\omega^2\,T}{2}}+
x_1''\sin{\frac{m\,\theta\,\omega^2\,T}{2}}\big)\Big)\sin{(y_2t)}
\Big)\, , \nn \\ \lb{e3.21}  \eea where $\csc u = \frac{1}{\sin
u}$. Inserting the above expressions and their time derivatives in
(\ref{e3.16}) we find \bea \nn L_{\theta}(\dot q,q) &=&
\frac{m\,\omega^2}{2\, \sin^2(\omega\, \sqrt{\kappa }\,\,T)}\\
\nn && \hp{-12mm} \times \Big(
\big(x_1''^2+x_2''^2\big)\cos(2\,\omega\,
\sqrt{\kappa }\,\,t ) +\big( x_1'^2+x_2'^2 \big) \\
\nn && \hp{-12mm} \times\cos(2\,\omega\, \sqrt{\kappa }\,\,(T-t)
)-2\,\cos(2\,\omega\, \sqrt{\kappa }\,\,(T-t) )\\
\nn && \hp{-12mm} \times\, \Big( \big( x_1'\,x_1''+
x_2'\,x_2'' \big) \cos{\frac{m\,\theta\,\omega^2\,T}{2}}\\
\label{e3.22} && \hp{-12mm}+\, \big( x_1''\,x_2'- x_1'\,x_2''\big)
\sin{\frac{m\,\theta\,\omega^2\,T}{2}} \Big) \Big) \,. \eea Using
(\ref{e3.22}),  we finally compute the corresponding action \bea
&& \nn  \hspace{-3mm } {\bar S}_\theta
(x'',T;x',0)=\int\limits_{0}^T \, L_{\theta}(\dot{q},q,t)\, d\,t
 = \! \nn \frac{m\,\omega}{2\,\sqrt{\kappa} \sin(\omega\,
\sqrt{\kappa }\,\,T)}  \\  \nn & & \hp{-2mm} \times
\Big(\big(x_1'^2+x_2'^2+x_1''^2+x_2''^2\big)\cos(\omega\,
\sqrt{\kappa}\,\,T)\\ \nn & & \hp{3mm} -\, 2\,\big(x_1'\,x_1''+
x_2'\,x_2''\big)\cos{\frac{m\,\theta\,\omega^2\,T}{2}}\\
\label{e3.23} & & \hp{3mm} +\, 2\,\big(x_1'\,x_2''-
x_2'\,x_1''\big)\sin{\frac{m\,\theta\,\omega^2\,T}{2}}\Big) \,.
\eea If we take into account expression (\ref{e3.1}) then firstly
we have\bea\nn \det{\left(-\frac{\partial^2 {\bar
S_{\theta}}}{\partial x''_k\,
\partial x'_j}\right)} =
\frac{m^2\,\omega^2}{\kappa \,
\sin^2(\omega\,\sqrt{\kappa}\,\,T)}\eea and finally \bea  \nn
 & & \hspace{-15mm} {\cal K}_\theta (x'',T;x',0) =\frac{1}{i\, h} \frac{m\,\omega}
 {\sqrt{\kappa}\, \big|\!\sin(\omega\,\sqrt{\kappa}\,\,T)\hp{-.1mm}\big|}
 \\ && \hp{0mm}\times  \exp\left(
 \frac{2\pi i}{h}\, \bar{S}_\theta (x'',T;x',0)   \right)
\label{e3.24},  \eea where $\bar{S}_\theta (x'',T;x',0)$ is given
by (\ref{e3.23}).

  Let us remark that in this case it is not so easy to find the
relation between the kernels of the harmonic oscillators in
commutative ($\theta=0$) and noncommutative regimes. Namely, in
commutative case ($\theta=0$ and $\kappa=1$), we have \bea && \nn
\hspace{-9mm } {\bar S}_0 (x'',T;x',0)=\int\limits_{0}^T \,
L_{0}(\dot{q},q,t)\, d\,t \vp{3mm} \\ & & \hp{-3mm} = \, \nn
\frac{m\,\omega}{2 \, \sin(\omega\,
T)}\, \Big(\big(x_1'^2+x_2'^2+x_1''^2+x_2''^2\big)\cos(\omega\, T)\\
 & & \hp{3mm} -\, 2\,\big(x_1'\,x_1''+ x_2'\,x_2''\big)\Big) \,,
\label{e3.25}
 \eea
and consequently
 \bea
 & & \hspace{-15mm} {\cal K}_0 (x'',T;x',0) =\frac{1}{i\,h}\, \frac{m\,\omega}
 {\big|\!\sin(\omega\,T)\hp{-.1mm}\big|} \nn \\ && \hp{12mm}\times  \exp\left(
 \frac{2\pi i}{h}\, \bar{S}_0 (x'',T;x',0)   \right).\label{e3.26}
  \eea

\section{Concluding Remarks}

Using transformation (\ref{e2.11}) for noncommuting coordinates
$\hat{x}_k$ in the quadratic Hamiltonian (\ref{e2.8}) we again
obtain quadratic Hamiltonian (\ref{e2.12}) but with commutative
spatial coordinates $\hat{q}_k$. To employ Feynman's path integral
we derived the corresponding effective quadratic Lagrangian
(\ref{e2.16}) and connection of coefficients (\ref{e2.17}) in
noncommutative regime $(\theta \ne 0)$ with those in the
commutative one $(\theta =0).$ The transition from Hamiltonian to
Lagrangian is performed using the usual formula $ L_\theta
(\dot{q},q,t) = \langle p,\dot{q} \rangle  - H_\theta (p,q,t)$. It
is worth noting that this transition can be also obtained as a
result of integration over momenta in the phase space path
integral (see Appendix and \cite{christiansen}). Since our
Lagrangian $L_\theta (\dot{q},q,t)$ is quadratic one can exploit
general expression for analytically evaluated Feynman's path
integral (\ref{e3.1}). Probability amplitudes on noncommutative
plane are calculated for a particle in a constant field
(\ref{e3.11}) and for a harmonic oscillator (\ref{e3.24}).

Note that the algebra of NCQM which is
\bea  && [\, \hat{x_a},\hat{p_b}\,] = i\,\hbar\,
(\delta_{ab} -\frac{1}{4} \, \theta_{ac}\, \sigma_{cb}), \ [\,\hat{x_a},\hat{x_b}\,] = i\,
\hbar \, \theta_{ab}, \nn \\
 && [\,\hat{p_a},\hat{p_b}\,] =i\, \hbar \, \sigma_{ab}, \ \ a,b,c =1,2,\dots, D   \label{e4.1} \eea
can be also transformed by linear transformation \bea \hat{ x_a} =
\hat{q_a} - \frac{\theta_{ab}\hat{k_b}}{2} ,\, \ \ \hat{ p_a} =
\hat{k_a} + \frac{\sigma_{ab}\hat{q_b}}{2} \label{e4.2} \eea to
the usual Heisenberg algebra (\ref{e1.1}) in terms of operators
$\hat{q_a},\, \hat{k_b}$. In the two-dimensional case  one has
$\theta_{ac} \, \sigma_{cb} = - \theta\, \sigma \, \delta_{ab}$
and it gives  $[\, \hat{x_a},\hat{p_b}\,] = i\,\hbar\, (1 + \frac{
\theta\, \sigma}{4})\, \delta_{ab}$ (see also (\cite{kochan})).
For phase space noncommutativity (\ref{e4.1}) and quadratic
Lagrangians one can perform a similar procedure described in the
preceding sections.

\begin{acknowledgments}

\noi This paper is completed during stay of B.D. in the Steklov
Mathematical Institute, Moscow. The work on this article was
partially supported by the Serbian Ministry of Science,
Technologies and Development under contracts No 1426 and No 1646.
The work of B.D. was also supported in part by RFFI grant
02-01-01084.
\end{acknowledgments}

\appendix

%\section{Appendixes}
\section{Path integral on a phase space for quadratic
Hamiltonians}

 We here show that a path integral on a phase space can be
reduced to the path integral on configuration space when
Hamiltonian $H(p,q,t)$ is any quadratic polynomial in $p$. Let us
start with path integral (\ref{e1.5}), where  $H(p,q,t)$ is given
by Eq. (\ref{e2.8}) or (\ref{e2.12}).

 By discretization of the time interval $t''-t'$ with equal
subintervals $ \epsilon=\frac{t''-t'}{N+1}$ and $t_n=t'+n\,
\epsilon,\, \, n=0,1, \cdots ,N+1 $ we have
$q(t_n)^{\tau}=q_n^{\tau}=(q_{n,1},q_{n,2},\cdots,  q_{n,D})$ and
$p(t_n)^{\tau}=p_n^{\tau}=(p_{n,1},p_{n,2},\cdots, p_{n,D})\, $ as
 position and momentum at the time $t_n\,.$ One can write
\bea & & \hspace{-3.5mm}{\cal K}(x'',t'';x',t') \label{e.a1}  \\
& & \hspace{-3mm} =\! \lim\limits_{N\longrightarrow\infty}
\!\int\limits_{\mathbb{R}^{DN}}\,
\int\limits_{\mathbb{R}^{D(N+1)}}\!\!\! \exp \left( \frac{2\pi
i}{h}\, S_N\right)\! \prod\limits_{n=1}^N d^D\,q_n \!
\prod\limits_{n=1}^{N+1}\frac{ d^{D}\,p_n}{h^{D}}\,,\nn \eea where
$S_N$ is \bea \label{e.a2} & & \nn\hspace{-12mm} S_N = \sum
_{n=1}^{N+1} ( \langle p_{n}, (q_{n}-q_{n-1}) \rangle - \epsilon (
\langle\, {A_n}\,{p_n} ,
{p_n}\, \rangle     \\
&  &  + \ \langle\, { B_n}\,{q_n} , {p_n}\, \rangle + \langle\,
{C_n}\,{q_n} , {q_n }\,\rangle + \langle\, {D_n} , {p_n} \,\rangle
\nn \\ & &  +\ \langle\, {E_n} , {q_n}\, \rangle +F_n ) )  . \eea
 $A_n,B_n,C_n,D_n,E_n $ and $F_n$ are the corresponding matrices
$(\det A_n \ne 0)$ and vectors which elements are taken at the
moment $t=t_n.$ Employing the Gauss integral \bea & &
\hspace{-12mm} \int\limits_{\mathbb{R}^{D}}\, \exp \left(
\frac{2\pi i}{h}\, [\, \langle u\, x, x\rangle + \langle
v,x\rangle \,] \right)\ d^D x \nn\\ & &  \hspace{-5mm} =
\frac{1}{\sqrt{(\frac{i}{h})^D\, \det (-2\, u)}}\, \exp \left(
\frac{-\,2\pi i}{4\,h}\,  \langle u^{-1}\, v, v \rangle \right) ,
\label{e.a3}\eea where $u$ is a nonsingular symmetric $D \times D$
matrix and $v$ is a $D$-dimensional vector, one can perform
integration over momenta $p_n$ and one obtains \bea &
&\hspace{-5mm} {\cal K}(x'',t'';x',t')  = \lim\limits_{N\to\infty}
\int\limits_{\mathbb{R}^{DN}}\, \prod\limits_{n=1}^N
\frac{d^D\,q_n}{\sqrt{(i\, h)^D\, \det (2\, \epsilon\hp{.2mm}
 A_n)}}\, \nn \\
& & \times \ \exp \Big( \frac{2\pi i\, \epsilon}{4\,h}\, \big[
\langle A_n^{-1}\,B_n\, q_n, B_n\,q_n\rangle - \langle A_n^{-1}\,
\dot{q_n}, B_n\,q_n\rangle \nn \\ & & - \ \langle A_n^{-1}\,B_n\,
q_n, \dot{q_n} \rangle + \langle A_n^{-1}\,\dot{q_n},
\dot{q_n}\rangle - 2\, \langle A_n^{-1}\,D_n, \dot{q_n}\rangle\nn \\
& & + \ 2\, \langle A_n^{-1}\,D_n, B_n\,q_n\rangle + \langle
A_n^{-1}\,D_n, D_n\rangle-\nn 4 \langle C_n\, q_n, q_n\rangle \\ &
& - \ 4\,  \langle E_n, q_n\rangle -4\, F_n  \big]
  \Big)\ . \label{e.a4}\eea

 Rewriting (\ref{e.a4})  in the form \bea & &\hspace{-5mm} {\cal
K}(x'',t'';x',t')  = \lim\limits_{N\to\infty}
\int\limits_{\mathbb{R}^{DN}}\, \prod\limits_{n=1}^{N+1}
\, \sqrt{\big(\frac{2}{i\, h\, \epsilon}\big)^D\, \det ( \alpha_n)}\, \nn \\
& & \times \ \exp \Big( \frac{2\pi i\, \epsilon}{h}\, \big[
\langle \al_n\, \dot{q_n}, \dot{q_n}\rangle + \langle \bet_n\,
q_n, \dot{q_n}\rangle + \langle \gamma_n\, q_n, q_n\rangle \nn
\\ & & + \ \langle \delta_n\, , \dot{q_n}, \rangle + \langle
\eta_n\, , q_n\rangle + \phi_n
  \big]\Big)\, \prod\limits_{n=1}^N
d^D\,q_n\,    , \label{e.a5}\eea and comparing with (\ref{e.a4})
we get \bea
  & & \hspace{-3mm} {\al_n } ={\frac 14}\,  {A_n}^{-1}, \hspace{2cm}   {\beta_n}
=-\,{1\over 2} \, {A_n}^{-1} {B_n}, \vp{2mm}\nn \\
& & \hspace{-3mm} {\gamma_n} ={1\over 4}\, {B_n}^{\tau} {A_n}^{-1} {B_n} -
{C_n } , \hspace{.1cm}
  {\delta_n} =- {1\over 2}\, {A_n }^{-1}  D_n,\vp{2mm} \label{e.a6}\\
& &  \hspace{-3mm} {\eta_n } = {1\over 2}\, {B_n}^{\tau} {A_n}^{-1}\,
{D_n}  - {E_n} , \hspace{.1cm} {\phi_n} = {1\over 4}\,
 {D_n}^{\tau}  {A_n}^{-1} {D_n}  - {F_n }
\nn \,. \eea Let us mention that the above formulas are the same
as the formulas (\ref{e2.9}) which are obtained on a different
way.

Now taking $N\to \infty$ one can rewrite (\ref{e.a5}) as
(\ref{e1.4})
\bea\label{e.a7} && \hspace{-25mm} {\cal K}(x'',t'';x',t') \nn \\
&& \hspace{-25mm} =\int \exp \Big( \frac{2\, \pi \, i}{h}\,
\int_{t'}^{t''} L (\dot{q}, q, t)\, d\,t \Big)\, {\cal D}q \, ,
\eea where the Lagrangian $L (\dot{q}, q, t)$ has the form
(\ref{e2.4}) and \bea && \hspace{-40mm} {\cal D}q  = \lim_{N \to
\infty} \prod\limits_{n=1}^N d\, \tilde{q}_n \, ,\quad     d\,
\tilde{q}_n = \left(\Big(\frac{2}{i\, h\, \epsilon}\Big)^D \, \det
(\alpha(t'')) \right)^{\frac{1}{ N}} \nn \\
\times \sqrt{\Big(\frac{2}{i\, h\, \epsilon}\Big)^D \, \det
(\alpha_n)}\, d\, q_n .
  \eea   Analytic evaluation of the path integral (\ref{e.a7})
yields  (\ref{1.6}) (see, e.g. \cite{steiner} and
\cite{grosjean}).

\newpage %Just because of unusual number of tables stacked at end
% \bibliography{apssamp}% Produces the bibliography via BibTeX.

%\end{document}

\end{document}